\documentclass[conference,a4paper]{IEEEtran}
\usepackage[utf8]{inputenc} 
\usepackage[T1]{fontenc}
\usepackage{url}              
\usepackage{cite}             

\usepackage[cmex10]{amsmath}  
\usepackage{mleftright}       
\mleftright                   

\usepackage{graphicx}         
\usepackage{booktabs}         

\usepackage{algorithm} 
\usepackage{algpseudocode}
\usepackage{algorithmicx}    
\usepackage{amsfonts}        
\usepackage{color}
\usepackage{caption}
\usepackage{subcaption}

\newtheorem{proposition}{Proposition}

\DeclareRobustCommand{\bm}[1]{\mathbf{\boldsymbol{#1}}}

\DeclareRobustCommand{\iter}[0]{T_{\rm it}}
\DeclareRobustCommand{\Poisson}[1]{{\rm Poisson}(#1)}
\newcommand{\1}{\mbox{1}\hspace{-0.25em}\mbox{l}}

\begin{document}

\title{Role of Bootstrap Averaging in Generalized Approximate Message Passing} 

\author{%
    \IEEEauthorblockN{Takashi Takahashi
                    }
    \IEEEauthorblockA{
                    Institute for Physics of Intelligence,
                    The University of Tokyo
                    \\
                    7-3-1 Hongo, Bunkyo-ku, Tokyo 113-0033, Japan,
                    takashi-takahashi@g.ecc.u-tokyo.ac.jp 
    }
}

\maketitle

\begin{abstract}
    Generalized approximate message passing (GAMP) is a computationally efficient algorithm for estimating an unknown signal \(\bm{w}_0\in\mathbb{R}^N\) from a random linear measurement \(\bm{y}=X\bm{w}_0 + \bm{\epsilon}\in\mathbb{R}^M\), where \(X\in\mathbb{R}^{M\times N}\) is a known measurement matrix and \(\bm{\epsilon}\) is the noise vector. The salient feature of GAMP is that it can provide an unbiased estimator \(\hat{\bm{r}}^{\rm G}\sim\mathcal{N}(\bm{w}_0, \hat{s}^2I_N)\), which can be used for various hypothesis-testing methods. In this study, we consider the bootstrap average of an unbiased estimator of GAMP for the elastic net. By numerically analyzing the state evolution of \emph{approximate message passing with resampling}, which has been proposed for computing bootstrap statistics of the elastic net estimator, we investigate when the bootstrap averaging reduces the variance of the unbiased estimator and the effect of optimizing the size of each bootstrap sample and hyperparameter of the elastic net regularization in the asymptotic setting \(M, N\to\infty, M/N\to\alpha\in(0,\infty)\). The results indicate that bootstrap averaging effectively reduces the variance of the unbiased estimator when the actual data generation process is inconsistent with the sparsity assumption of the regularization and the sample size is small. Furthermore, we find that when \(\bm{w}_0\) is less sparse, and the data size is small, the system undergoes a phase transition. The phase transition indicates the existence of the region where the ensemble average of unbiased estimators of GAMP for the elastic net norm minimization problem yields the unbiased estimator with the minimum variance.
\end{abstract}

\section{Introduction}
\label{sec:introduction}

Consider estimating an unknown signal $\bm{w}_0\in\mathbb{R}^N$ from a random linear measurement $\bm{y}\in\mathbb{R}^M$ in the form, 
\begin{equation}
    \bm{y} = X\bm{w}_0 + \bm{\epsilon}.
    \label{eq: measurement}
\end{equation}
In \eqref{eq: measurement}, $X\in\mathbb{R}^{M\times N}$ is a known measurement matrix whose elements are independent and identically distributed (i.i.d.) standard Gaussian variables, and $\bm{\epsilon}\sim\mathcal{N}(\bm{0},\Delta I_M)$ is the measurement noise. We also assume that each element of the unknown signal $\bm{w}_0$ is i.i.d., according to a distribution $q_0$.

Generalized approximate message passing (GAMP) \cite{rangan2011generalized,javanmard2013state} is a computationally efficient algorithm for solving this problem. A striking feature of GAMP is its applicability to various hypothesis testing \cite{sur2019modern}. Specifically, GAMP can provide an unbiased estimator $\hat{\bm{r}}^{\rm G}\sim \mathcal{N}(\bm{w}_0, \hat{s}^2I_M)$ \cite{javanmard2013state} in a high-dimensional asymptotic setting with $M,N\to\infty, M/N\to\alpha\in(0,\infty)$, where the variance $\hat{s}^2$ depends on the quality of the measurement $\bm{y}$ and denoising function used in GAMP. This unbiased estimator has been used to test the significance of estimated signals \cite{javanmard2014hypothesis, takahashi2018statistical, sur2019modern, sur2019likelihood}. 
 
The statistical power of these tests depends on the variance $\hat{s}^2$, with a lower variance leading to a higher statistical power. However, reducing the variance $\hat{s}^2$ is not a trivial task. Replacing the denoising function used in GAMP with a powerful one based on nonconvex regularization, for example, can worsen the convergence of GAMP \cite{sakata2021perfect} or, even if convergence is achieved, the improvement may be insignificant \cite{zheng2017does}. This study aims to find an alternative way to reduce the variance without nonconvex regularization.

To reduce the variance, we use the \emph{bootstrap averaging} \cite{efron1979bootstrap} of computational statistics (commonly known as \emph{ensemble learning} in machine learning \cite{zhou2012ensemble, murphy2022probabilistic}). Specifically, we consider averaging the unbiased estimators of GAMP for multiple bootstrap samples with arbitrary size $M\mu_B,\, \mu_B\in(0,\infty]$. For denoising function of GAMP, we consider the one for the elastic net \cite{zou2005regularization}. 

However, the efficient computation and theoretical analysis of an averaged unbiased estimator remains unresolved. To resolve this problem, we use \emph{AMP with resampling} (AMPR) \cite{obuchi2019semi, takahashi2020semi} that has been proposed for computing bootstrap statistics of the elastic net estimator by running a variant of GAMP once. We will argue that AMPR is actually computing the bootstrap average of the unbiased estimators of GAMP. That is, the averaged unbiased estimator can be computed efficiently, and its variance can be analyzed using the state evolution (SE) of AMPR, which has been developed to analyze the performance of AMPR in \cite{takahashi2020semi, obuchi2019semi}. We then conduct a thorough numerical analysis of the SE of the AMPR to investigate when bootstrap averaging reduces the variance of the unbiased estimator, and what phenomena occur when optimizing the bootstrap sample size and the hyperparameter of the elastic net regularization.

The findings of this study are summarized as follows:
\begin{itemize}
    \item As mentioned above, the averaged unbiased estimator is obtained by AMPR. Furthermore, its variance  can be estimated using the output of AMPR without knowing the actual signal $\bm{w}_0$. Thus, we can minimize the variance by adjusting the bootstrap sample size and the hyperparameters of the elastic net (see Section~\ref{sec: averaged unbiased estimator} and \ref{sec: variance optimization})
    \item The variance of the averaged unbiased estimator can be reduced via bootstrapping, especially when the true data generation process is inconsistent with the sparsity assumption of the regularization and the data size is insufficient (see Section~\ref{subsection: variance reduction})
    \item When \(\bm{w}_0\) is less sparse, and the data size is small, a phase transition occurs. This phase transition indicates the existence of the region where the value of the regularization parameter is infinitesimally small, and the number of unique data points in each bootstrap sample is less than the dimension of \(\bm{w}_0\). That is, in this region, the ensemble average of the unbiased estimators of GAMP for the elastic norm minimization problem (also known as the minimum norm \emph{interpolation} in machine learning \cite{muthukumar2020harmless, bartlett2020benign, hastie2022surprises}) yields the best averaged unbiased estimator (see Section~\ref{subsection: phase transition})
\end{itemize}


\subsection{Notation}
$\mathcal{N}(\mu, \sigma^2)$ denotes a Gaussian distribution with mean $\mu$ and variance $\sigma^2$ and ${\rm Poisson}(\mu_B)$ denotes a Poisson distribution with mean $\mu_B$. For a random variable $X\sim p_X$, we denote by $\mathbb{E}_X[\dots]$ an average $\int (\dots)p_X(x)dx$. Given a vector $\bm{x}\in\mathbb{R}^N$ and scalar function $f:\mathbb{R}\to\mathbb{R}$, we write $f(\bm{x})$ for the vector obtained by applying $f$ componentwise. For a vector $\bm{x}=(x_1,x_2,\dots,x_N)^\top\in\mathbb{R}^N$, we denote by $\bm{x}^2=(x_1^2,x_2^2,\dots,x_N^2)^\top$ the componentwise operations and by $\langle \bm{x}\rangle = N^{-1}\sum_{i=1}^Nx_i$ the empirical average. 

\section{Background on AMPR}
\label{sec: background on AMPR}

\begin{algorithm}[t]
    \caption{AMPR}
    \label{algo: AMPR}
    \begin{algorithmic}[1]
        \Require Measurement matrix $X\in\mathbb{R}^{M\times N}$, measurement $\bm{y}\in\mathbb{R}^M$, denoising function $g$ and its derivative $g^\prime$ from \eqref{eq: definition of denoising function} and \eqref{eq: definition of the derivative of denoising function}, the bootstrap sample size $\mu_B\in(0,\infty]$, and the number of iterations $T_{\rm it}$.
        \State Select initial $\bm{h}_0 \in\mathbb{R}^N, \hat{Q}_0, \hat{v}_0\in(0,\infty)$.
        \State Set $\bm{z}_0 =\bm{0}_M, \alpha_N=M/N$.
        \State Let $\bm{\xi}$ and $c$ be random variables distributed as $\bm{\eta}\sim\mathcal{N}(0,I_N)$ and $c\sim{\rm Poisson}(\mu_B)$.
        \For {$t=0,1,\dots,T_{\rm it}-1$}
            \State $\hat{\bm{w}}_t = \mathbb{E}_{\bm{\eta}}[g(\bm{h}_t + \sqrt{\hat{v}_t}\bm{\eta},\hat{Q}_t)]$.
            \State $\chi_t = \langle \mathbb{E}_{\bm{\eta}}[g^\prime(\bm{h}_t + \sqrt{\hat{v}_t}\bm{\eta}, \hat{Q}_t)] \rangle$.
            \State $v_t = \langle \mathbb{E}_{\bm{\eta}}[g(\bm{h}_t + \sqrt{\hat{v}_t}\bm{\eta}, \hat{Q}_t)^2] - \hat{\bm{w}}_t^2 \rangle $.
            \State $f^{(1)}_{t+1} = \mathbb{E}_c[\frac{c/\mu_B}{1 + c/\mu_B\chi_t}], \, f^{(2)}_{t+1}=\mathbb{E}_c[(\frac{c/\mu_B}{1 + c/\mu_B\chi_t})^2]$.
            \State $\hat{Q}_{t+1} = \alpha_N f^{(1)}_{t+1}$.
            \State $\bm{a}_{t+1} = f_{t+1}^{(1)}(\bm{y} - X\hat{\bm{w}_t} + \chi_t \bm{a}_t)$. \label{line: compute a}
            \State $\bm{h}_{t+1} = X^\top \bm{a}_{t} + \hat{Q}_{t+1}\hat{\bm{w}}_t$. \label{line: compute h}
            \State $\hat{v}_{t+1} = \alpha_N(f_{t+1}^{(2)}v_t + \frac{f_{t+1}^{(2)} - (f_{t+1}^{(1)})^2}{(f_{t+1}^{(t)})^2}\langle \bm{a}_{t+1}^2\rangle)$ .
        \EndFor
        \State \textbf{Return} $(\bm{h}_{\iter}, \hat{Q}_{\iter}, \hat{v}_{\iter})$.
    \end{algorithmic}
\end{algorithm}

\subsection{AMPR}
\label{subsec: AMPR}
Algorithm \ref{algo: AMPR} shows AMPR \cite{obuchi2019semi} with an elastic net denoising function. Function $g:\mathbb{R}\times(0,\infty)\to\mathbb{R}$ is the elastic net denoising function and $g^\prime$ is a derivative of $g$ with respect to the first argument: 
\begin{IEEEeqnarray}{rCl}
    g(h,\hat{Q}) &=& \left\{ \,
        \begin{IEEEeqnarraybox}[][c]{l?s}
            \IEEEstrut 
                0 & if $|h| \le \lambda\gamma $, \\ 
                \frac{h - {\rm sgn}(h)\gamma\lambda}{\hat{Q} + \lambda(1-\gamma)} & otherwise,
            \IEEEstrut 
        \end{IEEEeqnarraybox} 
    \right.
    \label{eq: definition of denoising function}
    \\
    g^\prime(h,\hat{Q}) &=& \left\{ \,
        \begin{IEEEeqnarraybox}[][c]{l?s}
            \IEEEstrut 
                0 & if $|h| \le \lambda\gamma $, \\ 
                \frac{1}{\hat{Q} + \lambda(1-\gamma)} & otherwise. 
            \IEEEstrut 
        \end{IEEEeqnarraybox} 
    \right.
    \label{eq: definition of the derivative of denoising function}
\end{IEEEeqnarray}
where $\lambda>0$ represents the regularization strength and $\gamma\in[0,1]$ is the $\ell_1$-ratio. At a fixed point, AMPR offers bootstrap statistics of the elastic net estimator as follows:

\begin{proposition}[bootstrap statistics based on AMPR \cite{obuchi2019semi}]
    \label{proposition: bootstrap statistics based on AMPR}
    Let $\hat{\bm{w}}^\ast$ be the elastic net estimator for a bootstrap sample $D^\ast$ of size $\mu_B M$.
    \begin{equation}
        \hat{\bm{w}}^\ast = \mathop{\rm argmin}_{\bm{w}\in\mathbb{R}^N}\sum_{\mu=1}^M \frac{c_\mu}{2\mu_B} (y_\mu - \bm{x}_\mu^\top\bm{w})^2 + \sum_{i=1}^N\lambda(\gamma|w_i| + \frac{1-\gamma}{2}w_i^2),
        \label{eq: elastic net single realization of c}
    \end{equation}
    where $c_\mu \sim_{\rm i.i.d.} {\rm Poisson}(\mu_B)$ represents the number of times the data point $(\bm{x}_\mu, y_\mu)$, $\bm{x}_\mu$ being the $\mu$-th row of $X$, appears in the bootstrap sample $D^\ast$. Then once the AMPR reaches its fixed point at sufficiently large $\iter$, the bootstrap statistics of $\hat{\bm{w}}^\ast$ can be computed as 
    \begin{multline}
        \mathbb{E}_{\bm{c}}\left[\psi(\hat{w}_i^\ast)\right] = \mathbb{E}_\eta[\psi(g(h_{i} + \sqrt{\hat{v}}\eta, \hat{Q})) ], 
        \quad \eta\sim\mathcal{N}(0,1),
        \label{eq: bootstrap statistics based on AMPR}
    \end{multline}
    where $\psi:\mathbb{R}\to\mathbb{R}$ is such that the expression in \eqref{eq: bootstrap statistics based on AMPR} is well-defined and otherwise arbitrary. The variables without iteration indexes $(h_i, \hat{v}, \hat{Q})$ are the output of AMPR at a fixed point.
\end{proposition}

 Note that the averages $\mathbb{E}_{\eta}[g(h+\sqrt{\hat{v}}\eta;\hat{Q})],\mathbb{E}_{\eta}[g(h+\sqrt{\hat{v}}\eta;\hat{Q})^2] $ and $\mathbb{E}_{\eta}[g^\prime (h+\sqrt{\hat{v}}\eta;\hat{Q})]$ can be written in closed-form by the error function, thus computing these quantities is computationally easy. Also, $\mathbb{E}_c[\dots]$ is an average over a one-dimensional discrete random variable. It can be computed numerically without much computational overhead. Hence the computational complexity of computing the RHS of \eqref{eq: bootstrap statistics based on AMPR} is dominated by the matrix-vector product operations in lines \ref{line: compute a}-\ref{line: compute h} of Algorithm \ref{algo: AMPR} instead of repeatedly computing $\hat{\bm{w}}^\ast$ for numerous realizations of $\bm{c}$, making AMPR a computationally efficient algorithm for computing the bootstrap statistics.

\subsection{SE of AMPR}
\label{subsec: SE of AMPR}

\begin{algorithm}[t]
    \caption{SE of AMPR}
    \label{algo: SE of AMPR}
    \begin{algorithmic}[1]
        \Require Initial state of AMPR $(\mathcal{E}_0, \hat{Q}_1, \hat{v}_1)$, variance of the measurement noise $\Delta$, distribution $q_0$ of the signal $\bm{w}_0$, measurement ratio $\alpha$, denoising function $g$ and its derivative $g^\prime$ from \eqref{eq: definition of denoising function} and \eqref{eq: definition of the derivative of denoising function}, the bootstrap sample size $\mu_B\in(0,\infty]$, and the number of iterations $T_{\rm it}$.
        \State Set $\hat{\chi}_1 = \alpha^{-1}\hat{Q}_1^2(\mathcal{E}_0 + \Delta)$
        \State Let $\xi, \eta, w_0$ and $c$ be random variables distributed as $\xi,\eta \sim \mathcal{N}(0,1), w_0\sim q_0$, and $c\sim{\rm Poisson}(\mu_B)$.
        \For {$t=1,\dots,T_{\rm it}-1$}
            \State $\hat{w}_t = g(\hat{Q}_t w_0 + \sqrt{\hat{\chi}_t} \xi + \sqrt{\hat{v}_t}\eta;\hat{Q}_t)$.
            \State $\mathcal{E}_t = \mathbb{E}_{w_0, \xi}[(\mathbb{E}_{\eta}[\hat{w}_t] - w_0)^2]$
            \State $\chi_t = \mathbb{E}_{w_0, \xi, \eta}[g^\prime(\hat{Q}_tw_0 + \sqrt{\hat{\chi}_t}\xi +  \sqrt{\hat{v}_t}\eta, \hat{Q}_t)]$.
            \State $v_t = \mathbb{E}_{w_0,\xi}[\mathbb{E}_\eta[\hat{w}_t^2] - \mathbb{E}_{\eta}[\hat{w}_t]^2 ] $.
            \State $f^{(1)}_{t+1} = \mathbb{E}_c[\frac{c/\mu_B}{1 + c/\mu_B\chi_t}], \, f^{(2)}_{t+1}=\mathbb{E}_c[(\frac{c/\mu_B}{1 + c/\mu_B\chi_t})^2]$.
            \State $\hat{Q}_{t+1} = \alpha f^{(1)}_{t+1}$.
            \State $\hat{\chi}_{t+1} = \alpha (f_{t+1}^{(1)})^2(\mathcal{E}_t + \Delta)$
            \State $\hat{v}_{t+1} = \alpha(f_{t+1}^{(2)}v_t + (f_{t+1}^{(2)} - (f_{t+1}^{(1)})^2)\mathcal{E}_t)$ .\label{line: se of AMPR vhat}
        \EndFor
    \end{algorithmic}
\end{algorithm}

AMPR displays remarkable behavior. Let $\hat{\bm{r}}_t = \bm{h}_t/\hat{Q}_t, t=1,2,\dots,\iter.$ Then $\hat{\bm{r}}_t$ behaves like a white Gaussian noise-corrupted version of the true signal $\bm{w}_0$ \cite{obuchi2019semi}. Furthermore, the variance can be estimated using SE.

\begin{proposition}[SE of AMPR \cite{obuchi2019semi}]
    \label{proposition: SE of AMPR}
    $\hat{\bm{r}}_t$ behaves as a white Gaussian noise-corrupted version of the actual signal $\bm{w}_0$:
    \begin{equation}
        \hat{\bm{r}}_t  \sim \mathcal{N}(\bm{w}_0, \hat{\sigma}_t^2), \quad \hat{\sigma}^2_t = \hat{\chi}_t/\hat{Q}_t^2,
        \label{eq: unbiased estimate of AMPR}
    \end{equation}
    for some positive value $\hat{\chi}_t$, indicating that $\hat{\bm{r}}_t$ can be used as an unbiased estimate of $\bm{w}_0$. The variance is predicted in the asymptotic setting $M, N\to\infty, M/N\to\alpha\in(0,\infty)$ using the scalar SE specified in Algorithm \ref{algo: SE of AMPR}. There, $\mathcal{E}_t, t=1,2,\dots$ corresponds to the mean squared error (MSE) of the AMPR estimate $\hat{\bm{w}}_t$: $\mathcal{E}_t = N^{-1}\|\hat{\bm{w}}_t - \bm{w}_0\|_2^2$.  To track the performance of AMPR, $\mathcal{E}_0$ should be inputted as the MSE of $\hat{\bm{w}}_0$.
\end{proposition}

Using the SE of the AMPR, we can predict the variance of the unbiased estimator in the asymptotic setting $M, N\to\infty, M/N\to\alpha\in(0,\infty)$ for each value of $\mu_B, \lambda$, and $\gamma$. Hence variance can be minimized by tuning $(\mu_B, \lambda, \gamma)$ using versatile black-box optimization methods implemented in various optimization libraries \cite{mogensen2018optim, 2020SciPy-NMeth}.

\section{Averaged unbiased estimator}
\label{sec: averaged unbiased estimator}
Here, we explain the meaning of the SE of AMPR. Subsequently, we argue that $\hat{\bm{r}}_t$ of AMPR is the bootstrap average of the unbiased estimators of GAMP.

The first point is the meaning of $\xi$ and $\eta$ that appear in SE. Propositions \ref{proposition: bootstrap statistics based on AMPR} and \ref{proposition: SE of AMPR} indicate that, once the AMPR reaches its fixed point at sufficiently large $\iter$, for any functions $\phi,\psi:\mathbb{R}\to\mathbb{R}$ such that the following expression is well defined, the following holds in the asymptotic setting $M, N\to\infty, M/N\to\alpha\in(0,\infty)$:
\vspace{-2truemm}
\begin{multline}
        \frac{1}{N}\sum_{i=1}^N \phi\left(
        \mathbb{E}_{\bm{c}}[\psi(\hat{w}_i^\ast)]
        \right) \to
        \mathbb{E}_{w_0,\xi}\left[
            \phi\left(
                \mathbb{E}_{\eta}[
                    \psi(g(\hat{Q}w_0 
                \right.
            \right.
                    \\
                \left.
            \left.
                    + \sqrt{\hat{\chi}}\xi + 
                    \sqrt{\hat{v}}\eta;\hat{Q}))
                ]
            \right)
        \right],
    \label{eq: decoupling principle}
\end{multline}
where the variables $\hat{Q}, \hat{\chi}, \hat{v}$ are the outputs of SE at a fixed point.
In \eqref{eq: decoupling principle}, one can interpret that $\xi$ and $\eta$ effectively represents the randomness originating from $(X,\bm{y})$ and $\bm{c}$, respectively.

The next point is the relationship between AMPR and GAMP. For this, we define $\bar{q}_t \equiv \mathcal{E}_t + v_t$ and $\bar{\chi}_t = \hat{\chi}_t + \hat{v}_t$. Then, the evolution of $(\hat{Q}_t, \bar{\chi}_t, \bar{q}_t, \chi_t)$ is equivalent to the SE of GAMP for computing $\hat{\bm{w}}^\ast$ for one realization of $\bm{c}$ \cite{rangan2011generalized, javanmard2013state}. Recall that $\hat{\bm{w}}^\ast$ is specified in \eqref{eq: elastic net single realization of c}. In addition, $\bar{\chi}_t/\hat{Q}_t^2\,(=(\mathcal{E}_t + \hat{v}_t)/\hat{Q}_t^2)$ represents the variance of the unbiased estimate of GAMP $\hat{\bm{r}}_t^{\rm G}(\bm{c})$ at each iteration.  These observations indicate that the unbiased estimator of GAMP at each iteration $t$ can be decomposed into two Gaussian variables:
\begin{IEEEeqnarray}{rCl}
    \hat{\bm{r}}_t^{\rm G}(\bm{c}) &=& \bm{w}_0 + \frac{\sqrt{\bar{\chi}_t}}{\hat{Q}_t}\bm{\xi_0} = \bm{w}_0  + \frac{\sqrt{\hat{\chi}_t}}{\hat{Q}_t}\bm{\xi} + \frac{\sqrt{\hat{v}_t}}{\hat{Q}_t}\bm{\eta},
    \label{eq: decomposition of noise}
\end{IEEEeqnarray}
where $\bm{\xi}_0,\bm{\xi},\bm{\eta}\sim\mathcal{N}(0,I_N)$, and $\bm{\xi}$ and $\bm{\eta}$ represent the randomness originating from $(X,\bm{y})$ and $\bm{c}$, respectively. We can interpret that AMPR computes the bootstrap statistics by decomposing $\bm{\xi}_0$ into $\bm{\xi}$ and $\bm{\eta}$, and averaging over $\bm{\eta}$. This yields the unbiased estimator $\hat{\bm{r}}_t$ as in \eqref{eq: unbiased estimate of AMPR} that is the bootstrap average of $\hat{\bm{r}}^{\rm G}(\bm{cka})$. We verify this point numerically in section \ref{subsec: distribution of the unbiased estimator} (see Fig. \ref{fig: distribution of the unbiased estimator, single-shot case} and \ref{fig: distribution of the unbiased estimator, averaged case}).

    \begin{figure}[t]
        \centering
        \includegraphics[width=0.9\linewidth]{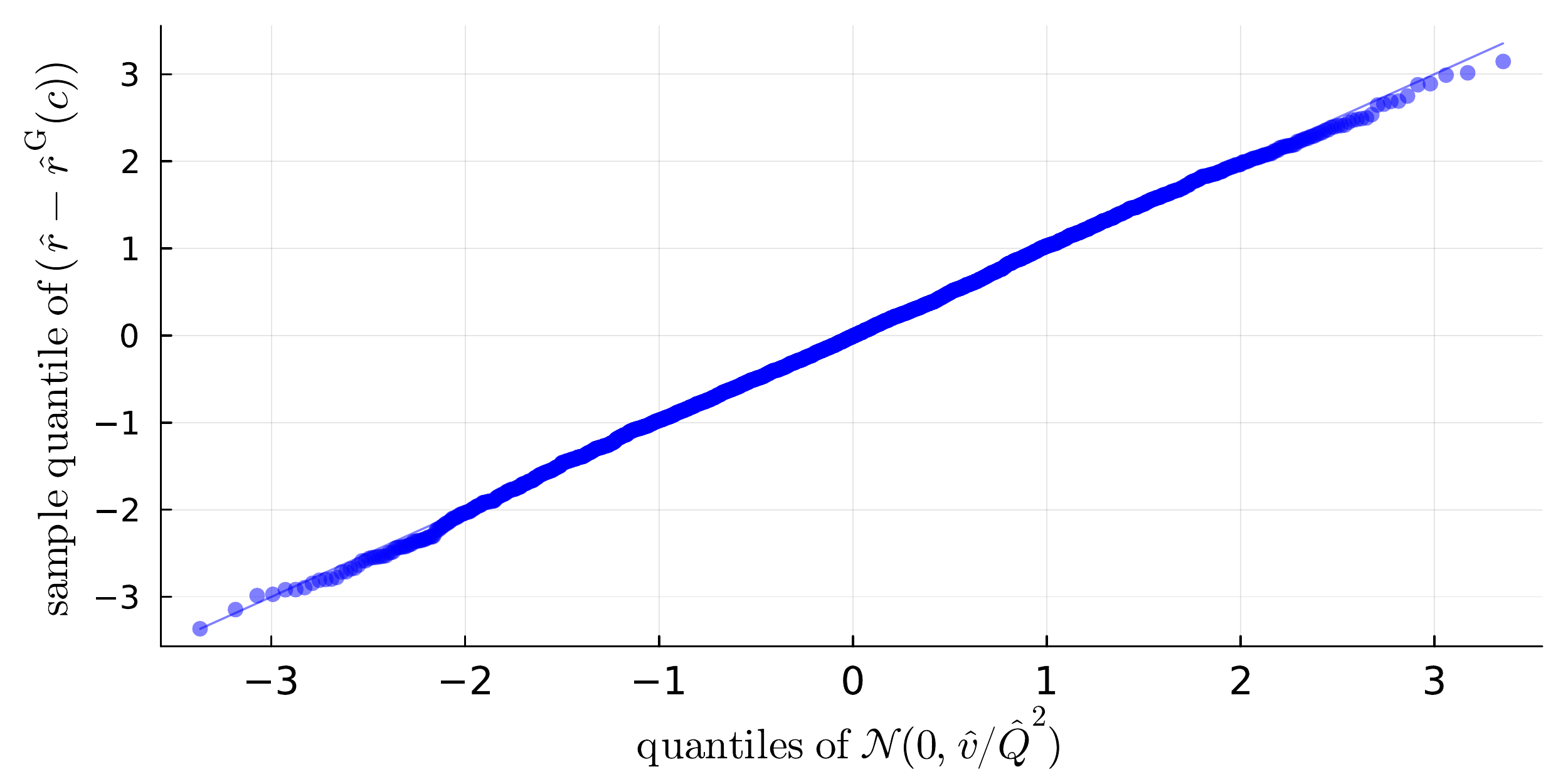}
        \caption{Q-Q plot of $\hat{\bm{r}} - \hat{\bm{r}}^{\rm G}(\bm{c})$ for one realization of $X,\bm{y}$, and $\bm{c}$. The parameters are set as $(N,\alpha, \Delta, \lambda, \gamma, \mu_B, \rho)=(4096, 0.25, 0.1, 0.5, 0.5, 0.1)$.}
        \label{fig: distribution of the unbiased estimator, single-shot case}
    \end{figure}

    \begin{figure}[t]
        \centering
        \includegraphics[width=0.9\linewidth]{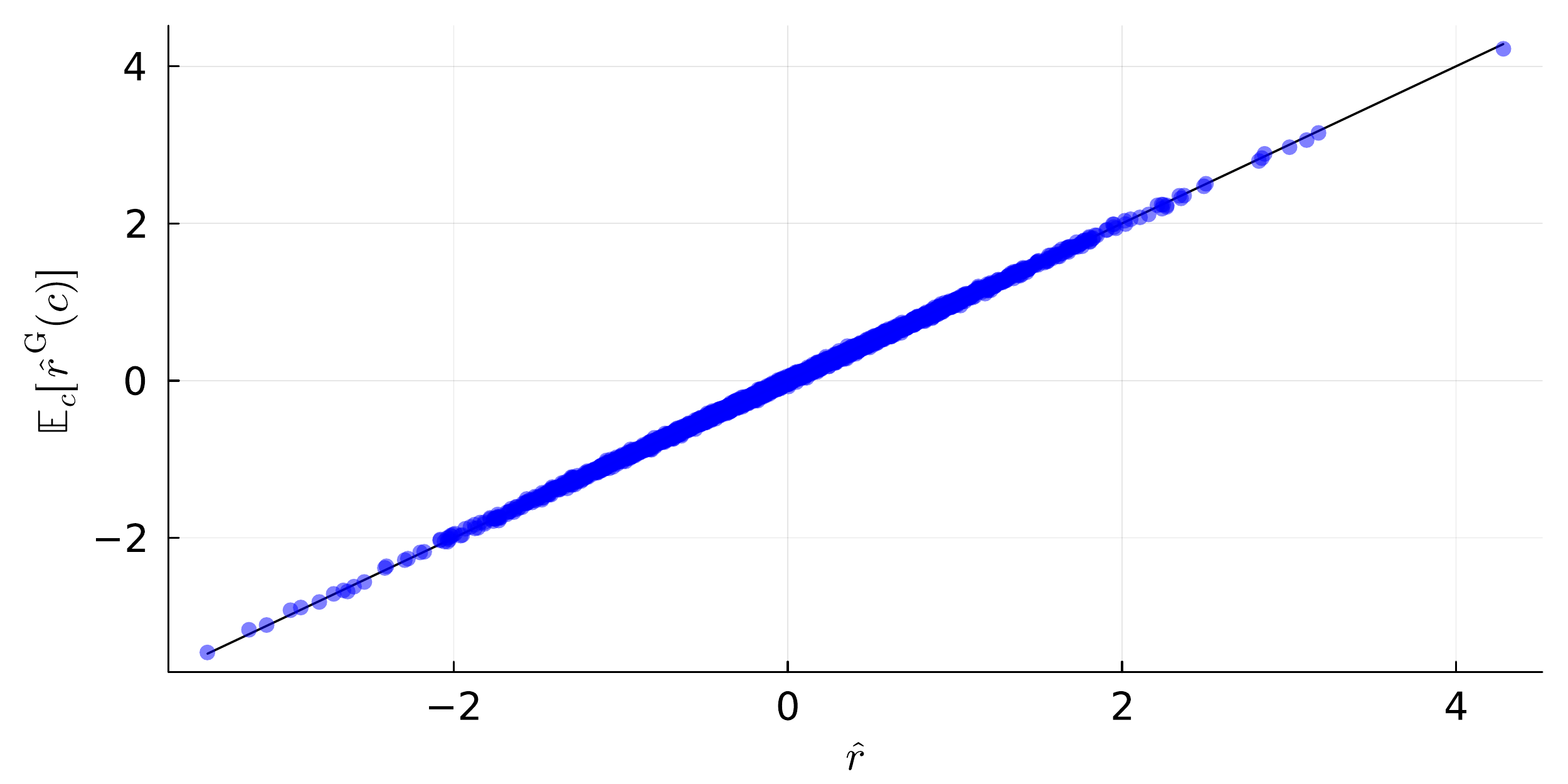}
        \caption{Scatter plot for $\hat{\bm{r}}$ and $\mathbb{E}_{\bm{c}}[\hat{\bm{r}}^{\rm G}(\bm{c})]$ for one realization of $X,\bm{y}$. The parameters are set as $(N,\alpha, \Delta, \lambda, \gamma, \mu_B, \rho)=(4096, 0.8, 0.25, 0.1, 0.5, 0.5, 0.1)$. $\mathbb{E}_{\bm{c}}[\hat{\bm{r}}^{\rm G}(\bm{c})]$ was computed from $65536$ realizations of $\bm{c}$.}
        \label{fig: distribution of the unbiased estimator, averaged case}
    \end{figure}

    \begin{figure*}[t]
        \centering
        \subfloat[$\ell_1$-ratio is fixed: $\gamma=1$.]{%
            \label{fig: subfig variance reduction l1ratio fixed 1.0}
            \includegraphics[width=5.15cm]{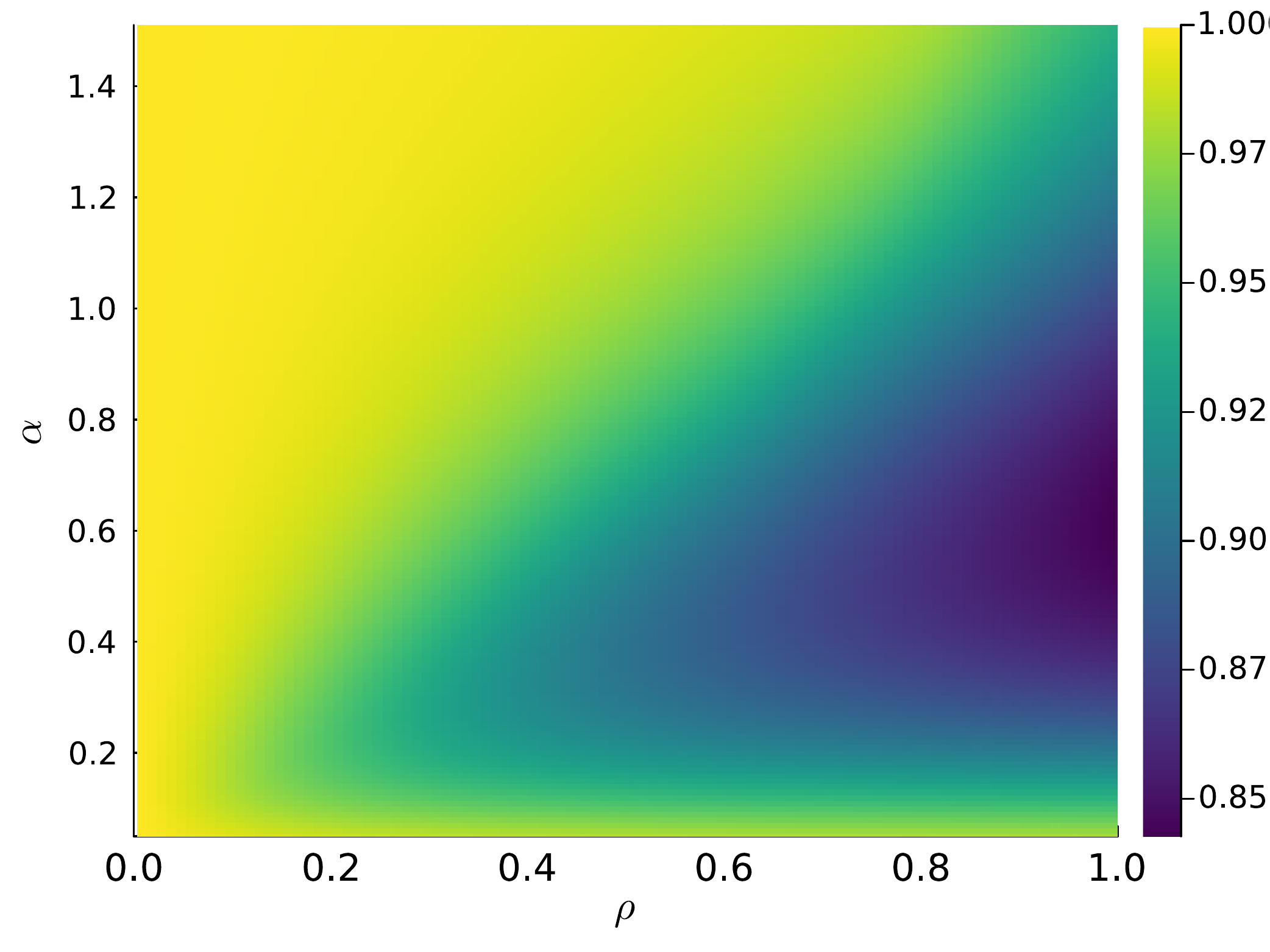}
        }
        \hfil
        \subfloat[$\ell_1$-ratio is fixed: $\gamma=1/2$.]{%
            \label{fig: subfig variance reduction l1ratio fixed 0.5}
            \includegraphics[width=5.15cm]{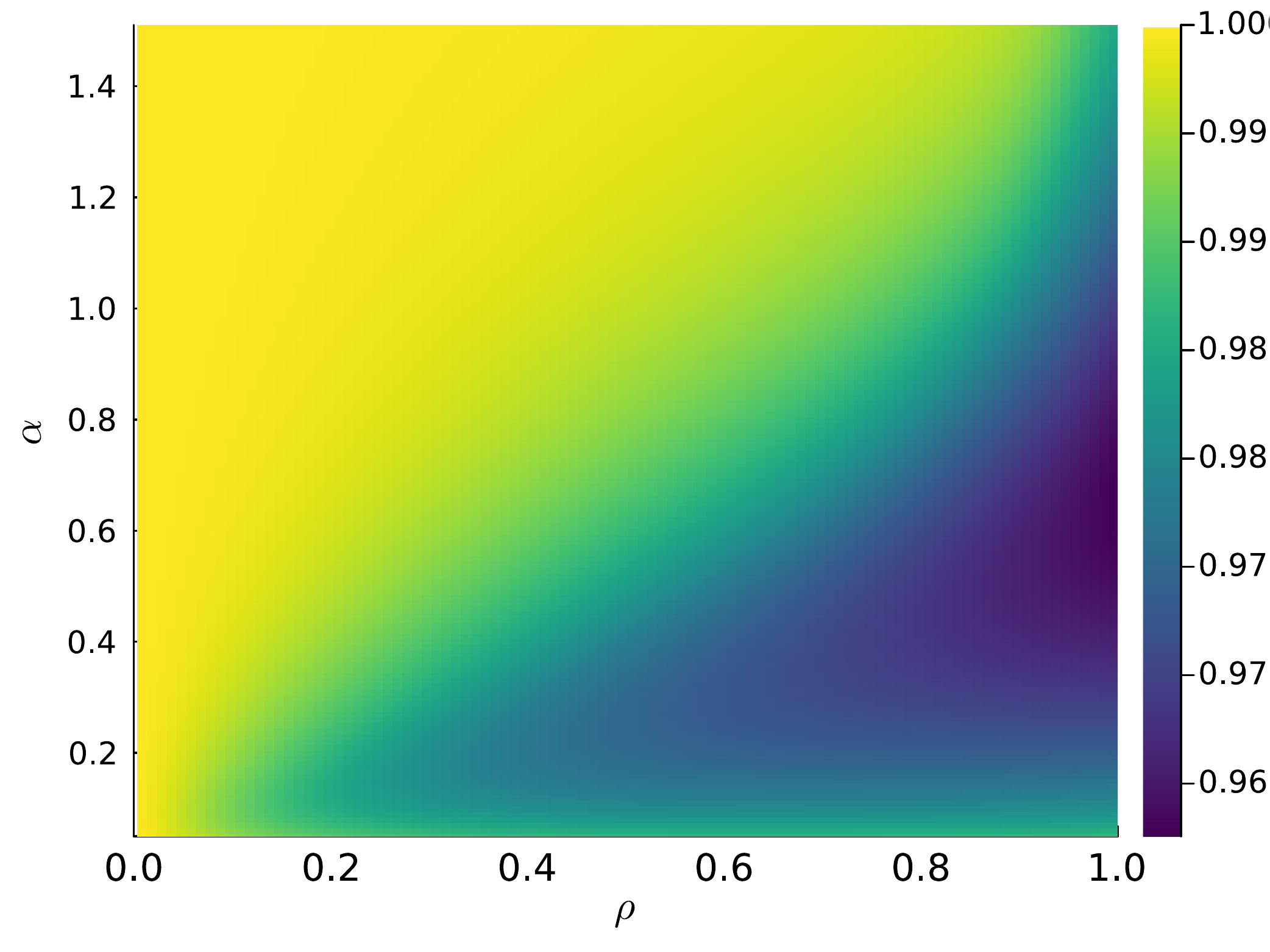}
        }
        \hfil
        \subfloat[Optimal $\ell_1$-ratio.]{%
            \label{fig: subfig variance reduction elastic net}
            \includegraphics[width=5.15cm]{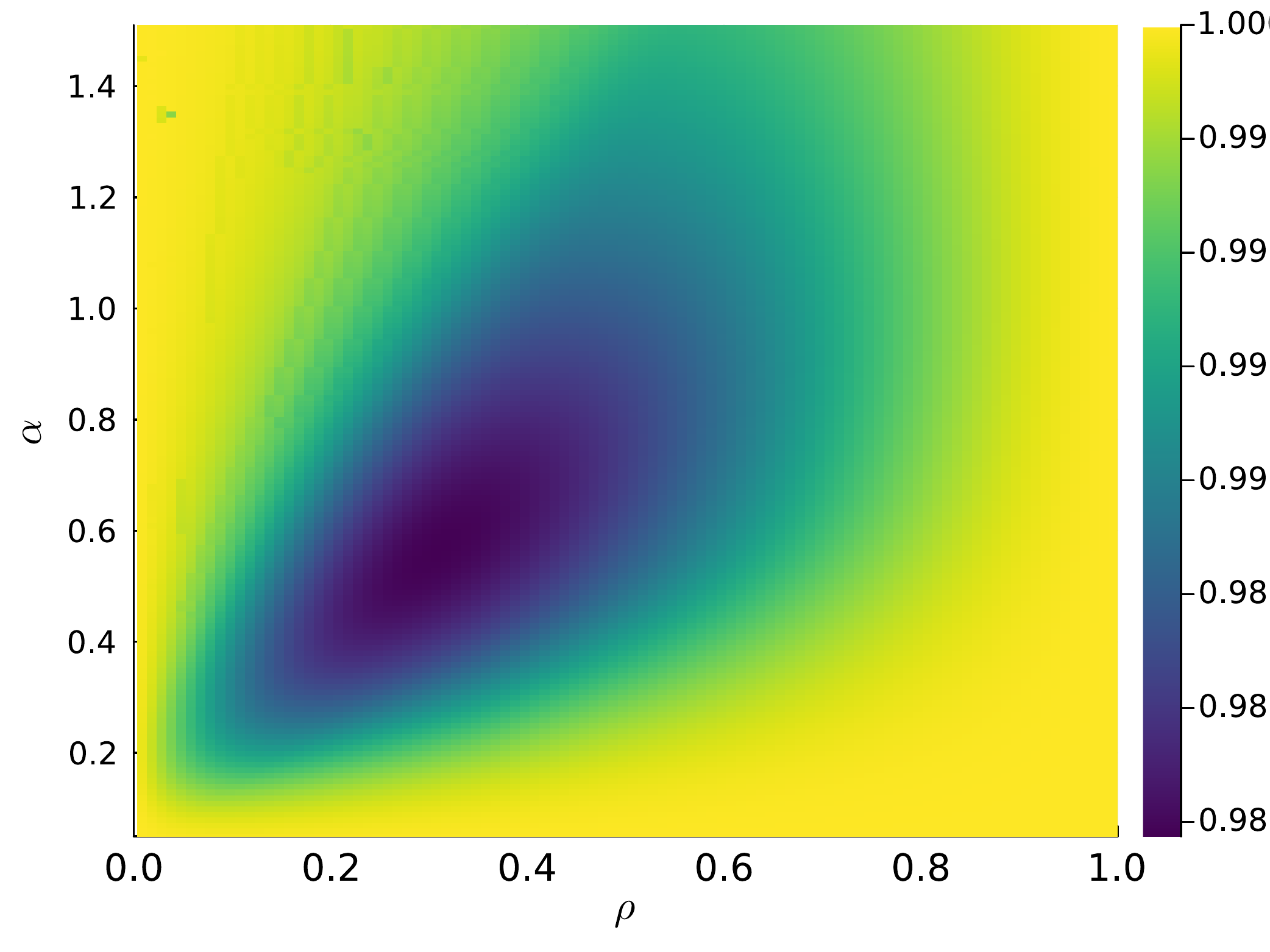}
        }
        \caption{
            Ratio of optimal variances $\hat{\sigma}^2/\hat{s}^2$ is plotted against the sparsity of the true signal $\rho$ and the measurement ratio $\alpha$ as heat maps. In panels (a) and (b), the $\ell_1$-ratio of the elastic net regularization is fixed. In panel (c), the $\ell_1$-ratio is also optimized. The measurement noise is set as $\Delta=0.15$.
        }
        \label{fig: ratio of optimal variances}
    \end{figure*}

\section{Variance optimization}
\label{sec: variance optimization}
In this section, we describe the properties of the variance of the unbiased estimate of AMPR.

First, the following proposition states that bootstrap averaging for the optimal choice of $(\mu_B, \lambda, \gamma)$ does not increase the variance of the unbiased estimator.
\begin{proposition}
    \label{proposition: variance should be smaller}
    Let $\hat{s}^2$ be the variance of the unbiased estimator of GAMP at a fixed point to compute the elastic net estimator $\hat{\bm{w}}$ of $D=(X, \bm{y})$: $\hat{\bm{w}} = \mathop{\rm argmin}_{\bm{w}\in\mathbb{R}^N}\frac{1}{2}\sum_{\mu=1}^M (y_\mu - \bm{a}_\mu^\top \bm{w})^2 + \sum_{i=1}^N\lambda(\gamma|w_i| + \frac{1-\gamma}{2})w_i^2$.
    We can select the parameter $(\mu_B^\star, \lambda^\star, \gamma^\star)$ such that the variance of AMPR's unbiased estimator $\hat{\sigma}^2$ at a fixed point does not exceed $\hat{s}^2$.
\end{proposition}
\begin{IEEEproof}
    We denote by $\hat{v}, v$ the values of $\hat{v}_t$ and $v_t$ at the fixed point of the SE of AMPR. We consider the limit $\mu_B\to\infty$. In this limit, the random variable $c/\mu_B, c\sim\Poisson{\mu_B}$ that appears in \eqref{eq: elastic net single realization of c} and Algorithm \ref{algo: AMPR} behaves deterministically; the mean and variance of $c/\mu_B$ converge to $1$ and $0$. Then proposition \ref{proposition: bootstrap statistics based on AMPR} indicates that $\hat{v}=0$. Moreover, from line \ref{line: se of AMPR vhat} in Algorithm \ref{algo: SE of AMPR}, $\hat{v}=0$ implies $v=0$, which yields the GAMP algorithm for computing the elastic net estimator of $D$ \cite{rangan2011generalized}.
\end{IEEEproof}

Although SE prediction of the variance of $\hat{\bm{r}}_t$ requires information on the unknown signal $\bm{w}_0$, we can predict the variance from the data. In other words, estimating the variance of $\hat{\bm{r}}_t$ does not require explicit knowledge on $\bm{w}_0$.
\begin{proposition}[Variance estimation from data]
    \label{proposition: variance estimation from data}
    In the asymptotic setting $M,N\to\infty, M/N\to\alpha\in(0,\infty)$, the variance $\hat{\sigma}_t^2$ of the unbiased estimate $\hat{\bm{r}}_{t}$ can be estimated as 
    \begin{equation}
        \hat{\sigma}_t^2 = \alpha \langle \bm{a}_t^2 \rangle /\hat{Q}_t^2.
    \end{equation}
\end{proposition}
\noindent\begin{IEEEproof}
    In SE of AMPR, $\hat{\chi}_t/\hat{Q}_t^2$ is determined by the MSE $\mathcal{E}_{t-1}$ and variance of measurement noise $\Delta$ as $\hat{\chi}_{t}/\hat{Q}_{t}^2 = \alpha^{-1}(\mathcal{E}_{t-1} + \Delta)$. For linear models, $\mathcal{E}_{t-1} + \Delta$ corresponds to the prediction error for a new sample and can be estimated using the leave-one-out error (LOOE) \cite{obuchi2016cross}. LOOE can be estimated from $\bm{a}_{t}$ because it is proportional to the leave-one-out estimate for the data point $\mu$: $a_{t,\mu} = \hat{Q}_{t}\alpha^{-1}(y_\mu - \bm{x}_\mu^\top \hat{\bm{w}}_t^{\backslash \mu})$, where $\hat{\bm{w}}_t^{\backslash\mu}$ is the AMPR's estimate of $\bm{w}_0$ without the sample $\mu$ (Equation (19) of reference \cite{obuchi2019semi}).
\end{IEEEproof}

Propositions \ref{proposition: variance should be smaller} and \ref{proposition: variance estimation from data} indicate that the variance $\hat{\sigma}^2$ can be minimized even if the signal $\bm{w}_0$ is unknown. However, we use SE for the theoretical assessment in the next section for convenience.

\section{Numerical analysis}
\label{sec: SE analysis} 
In the sequel, by numerically minimizing the variance using SE, we investigate when bootstrapping reduces the variance of the unbiased estimator and the phenomena that occur when optimizing the bootstrap sample size $\mu_B$ and the hyperparameter of the elastic net regularization $(\lambda, \gamma)$.  

For this, we searched for the optimal parameter $(\mu_B^\star, \lambda^\star, \gamma^\star)$ that yielded the minimum variance using the SE of AMPR and the Nelder-Mead algorithm in the \emph{Optim.jl} library \cite{mogensen2018optim}. We obtained the fixed point of the SE by iterating the SE a sufficient number of times. For comparison, the same optimization was performed for the non-bootstrap case. For the signal distribution $q_0$, we consider the Gauss-Bernoulli model: $q_0 = \rho \delta_0 + (1-\rho)\mathcal{N}(0,1)$, with sparsity $\rho\in(0,1)$.

In this section, we denote the outputs of the AMPR or GAMP at fixed points by unindexed variables.

\subsection{Distribution of the unbiased estimator}
\label{subsec: distribution of the unbiased estimator}
We verified the interpretation of AMPR's output $\hat{\bm{r}}$ described in Section \ref{sec: averaged unbiased estimator}.  For this, we compared the output of GAMP $\hat{\bm{r}}^{\rm G}(\bm{c})$ for each realization of $\bm{c}$ as in \eqref{eq: elastic net single realization of c}, and the output of AMPR. The parameters used to produce the figure were set as  $(N,\alpha, \Delta, \lambda, \gamma, \mu_B, \rho)=(4096, 0.8, 0.25, 0.1, 0.5, 0.5, 0.1)$.

Fig. \ref{fig: distribution of the unbiased estimator, single-shot case} shows the sample quantiles of $\hat{\bm{r}} - \hat{\bm{r}}^{\rm G}(\bm{c})$ versus the normal distribution with variance $\hat{v}/\hat{Q}^2$. The scattered points are approximately aligned with a line with a slope of $1$ and an intercept of $0$. Fig. \ref{fig: distribution of the unbiased estimator, averaged case} shows the scatter plot of $\hat{\bm{r}}$ versus $\mathbb{E}_{\bm{c}}[\hat{\bm{r}}^{\rm G}(\bm{c})]$. Again, the scattered points are approximately aligned with a line with a slope of $1$ and an intercept of $0$. This is consistent with the decomposition of the Gaussian noise \eqref{eq: decomposition of noise}. Thus, AMPR computes the bootstrap average of the unbiased estimator of GAMP.

\subsection{Variance reduction}
\label{subsection: variance reduction}

We quantitatively compare the variance $\hat{\sigma}^2$ of the averaged unbiased estimator with that of $\hat{s}^2$ without bootstrapping. Fig. \ref{fig: ratio of optimal variances} shows the ratio of the optimal $\hat{\sigma}^2$ and the optimal $\hat{s}^2$. Panels (a) and (b) show the results when the $\ell_1$-ratio $\gamma$ is fixed and panel (c) shows when the $\ell_1$-ratio is optimized. As expected from Proposition \ref{proposition: variance should be smaller}, the variance is reduced compared to the case without bootstrapping. The magnitude of the reduction is larger when the $\ell_1$-ratio is fixed and close to $1$ (LASSO \cite{tibshirani2011regression} case). In particular, the largest improvement is obtained when $\rho$ is large (less sparse) and the measurement ratio $\alpha$ is small. The improvement is minor when $\ell_1$-ratio is optimized. This suggests that using the bootstrap average is effective when the actual data generation process is inconsistent with the sparsity assumption of the regularization and the data size is insufficient. However, when the data size is too small, meaningful improvement cannot be obtained.


    \begin{figure*}[t]
        \centering
        \subfloat[$\ell_1$-ratio is fixed: $\gamma=1$]{%
            \label{fig:subfigA}
            \includegraphics[width=5.05cm]{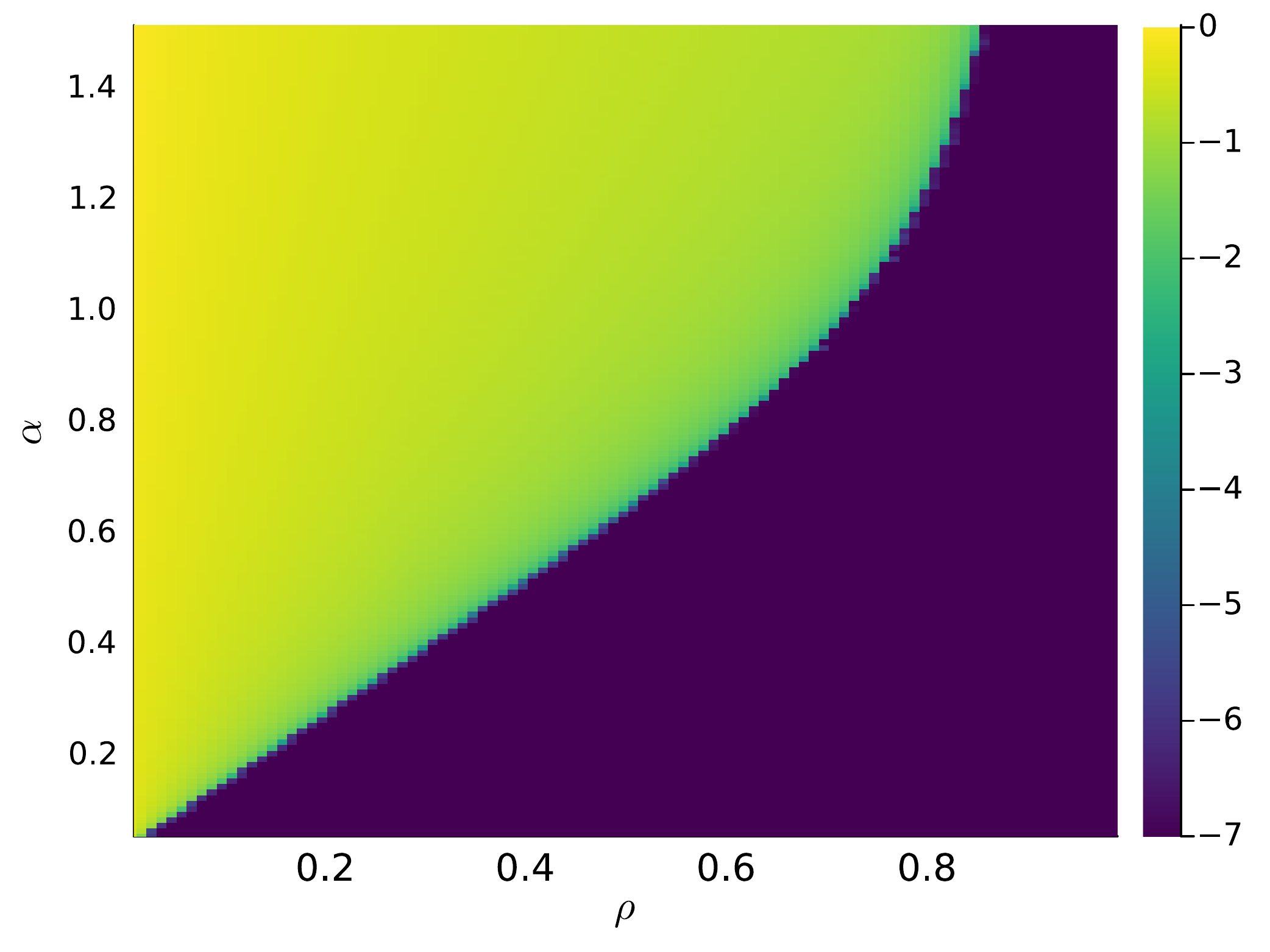}
        }
        \hfil
        \subfloat[$\ell_1$-ratio is fixed: $\gamma=1/2$.]{%
            \label{fig:subfigB}
            \includegraphics[width=5.05cm]{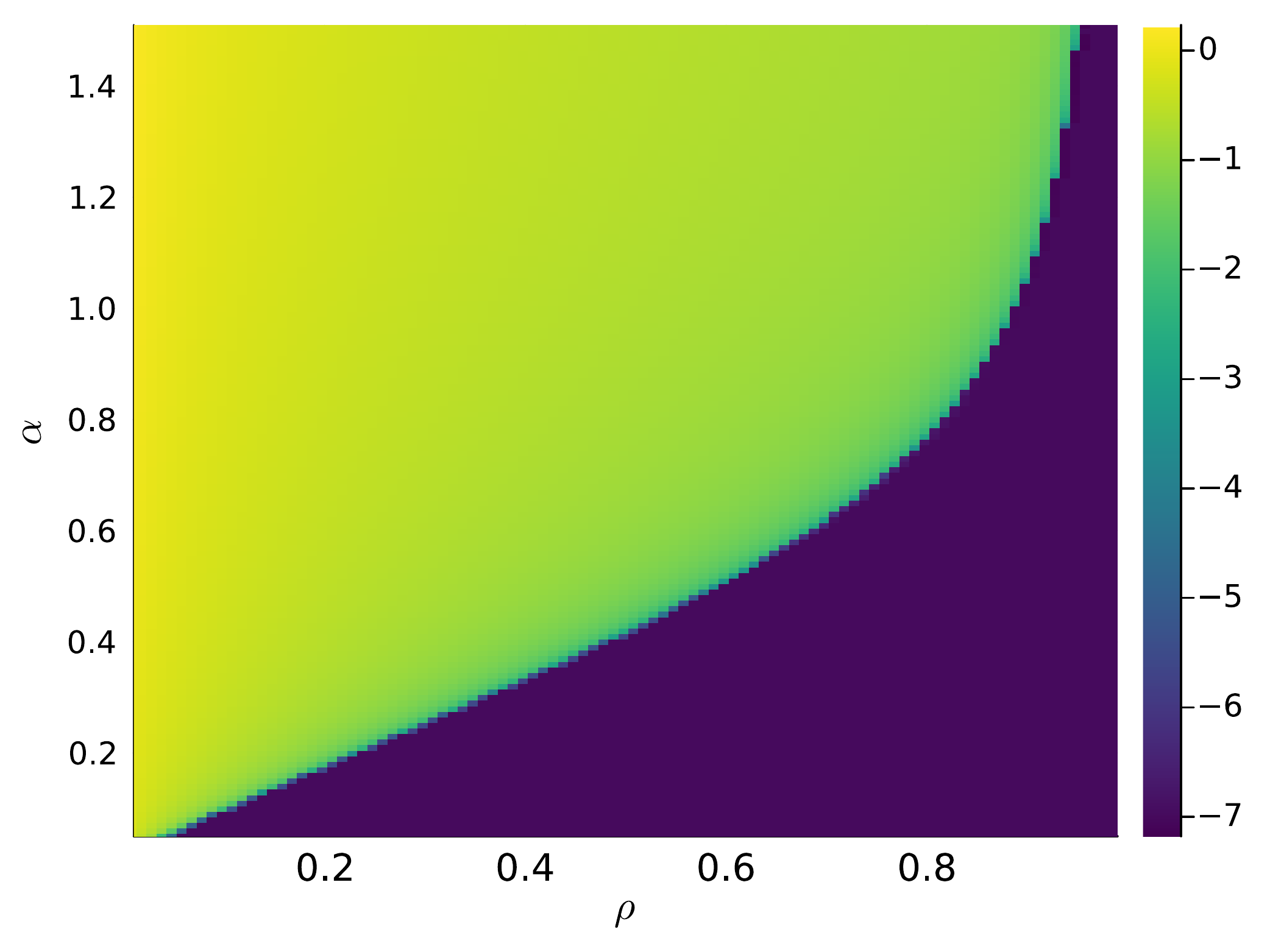}
        }
        \hfil
        \subfloat[Optimal $\ell_1$-ratio.]{%
            \label{fig:subfigB}
            \includegraphics[width=5.05cm]{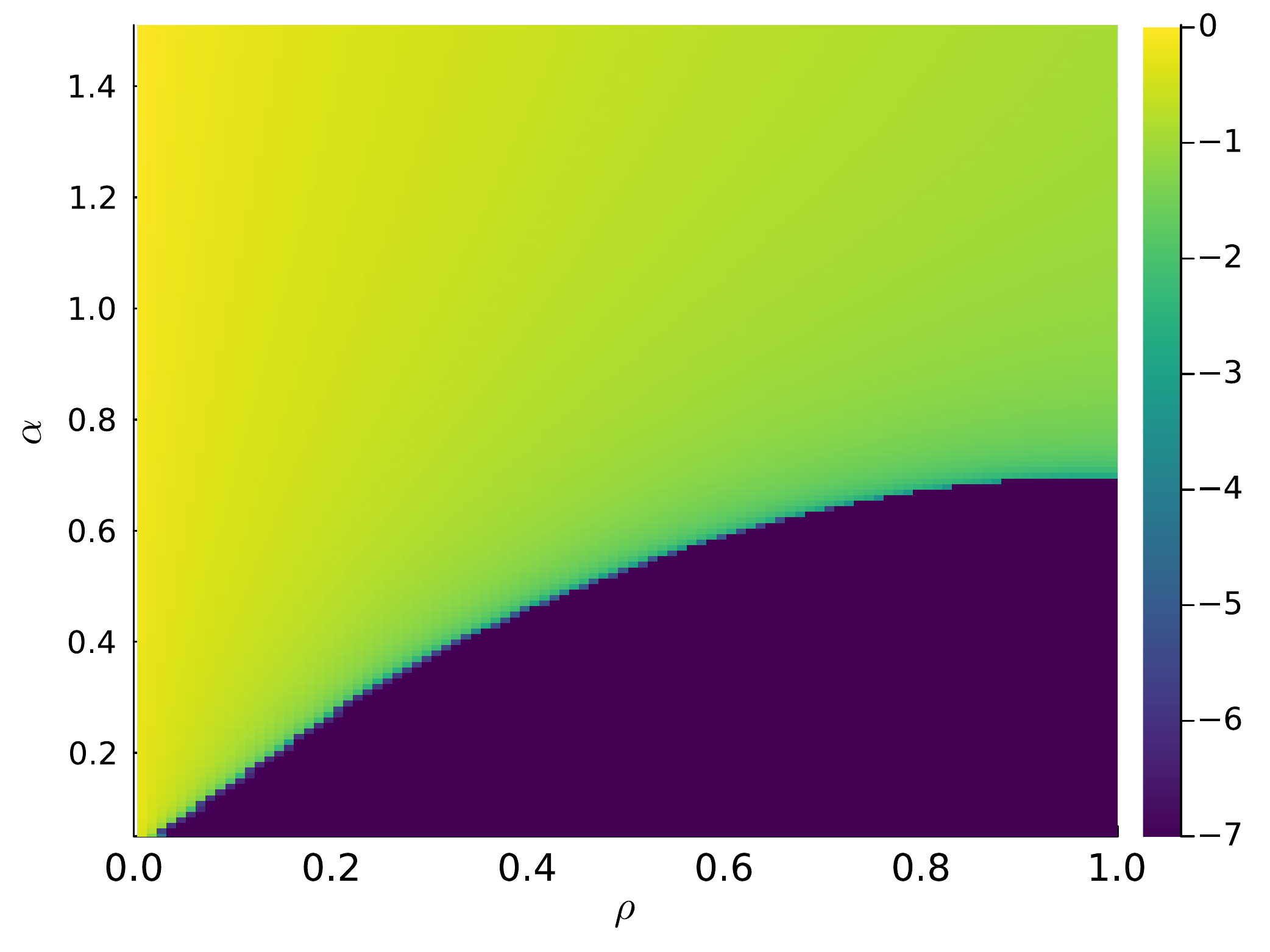}
        }
        \caption{The optimal regularization parameter $\log_{10} \lambda$ for the bootstrap averaged unbiased estimator are plotted against the sparsity of the true signal $\rho$ and the measurement ratio $\alpha=M/N$ as a heat map. The measurement noise is set as $\Delta=0.15$.}
        \label{fig: optimal regularization}
    \end{figure*}
    \vspace{-1truemm}

    \begin{figure*}[t]
        \centering
        \subfloat[$\ell_1$-ratio is fixed: $\gamma=1$]{%
            \label{fig:subfigA}
            \includegraphics[width=4.8cm]{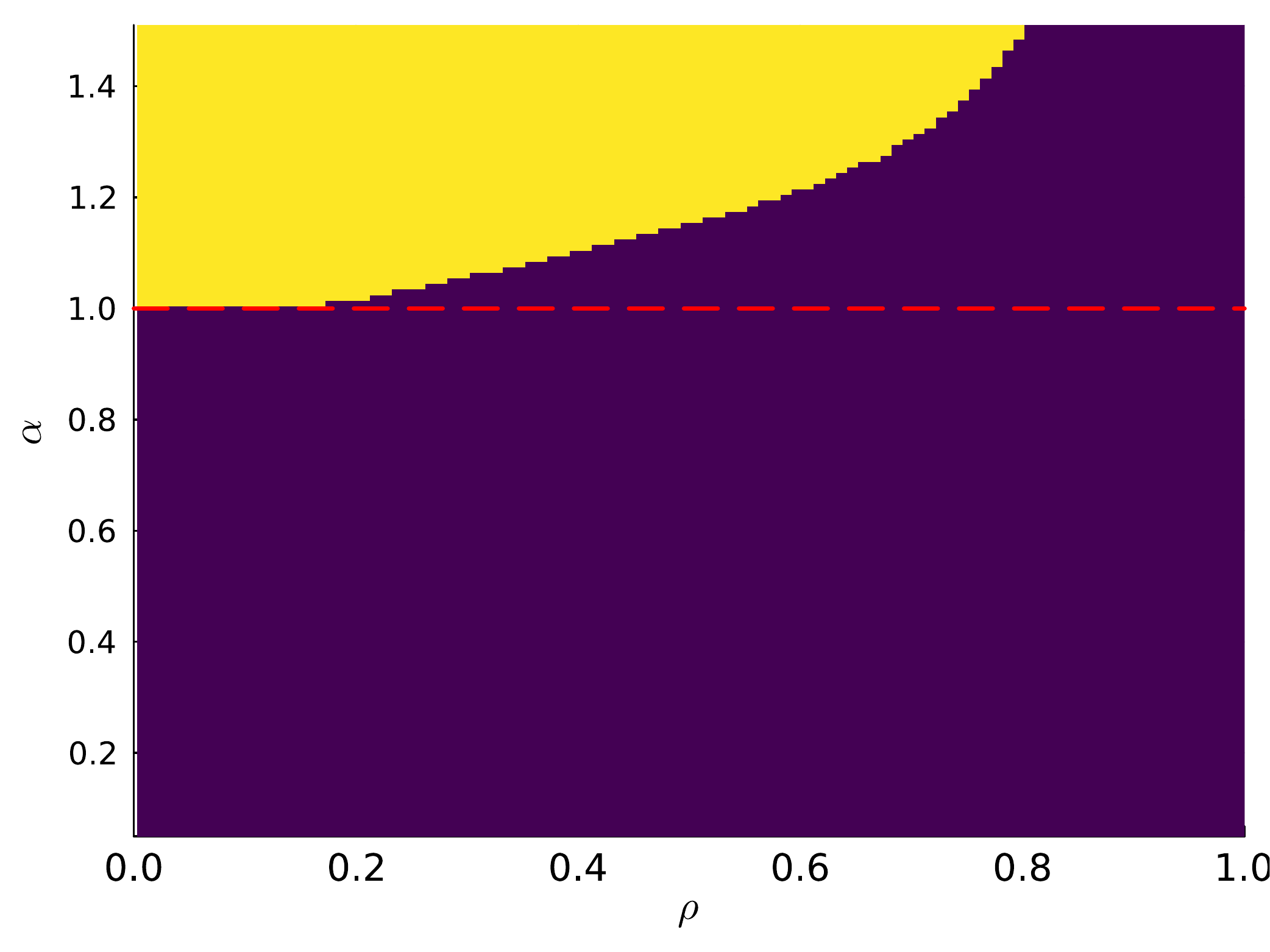}
        }
        \hfil
        \subfloat[$\ell_1$-ratio is fixed: $\gamma=1/2$.]{%
            \label{fig:subfigB}
            \includegraphics[width=4.8cm]{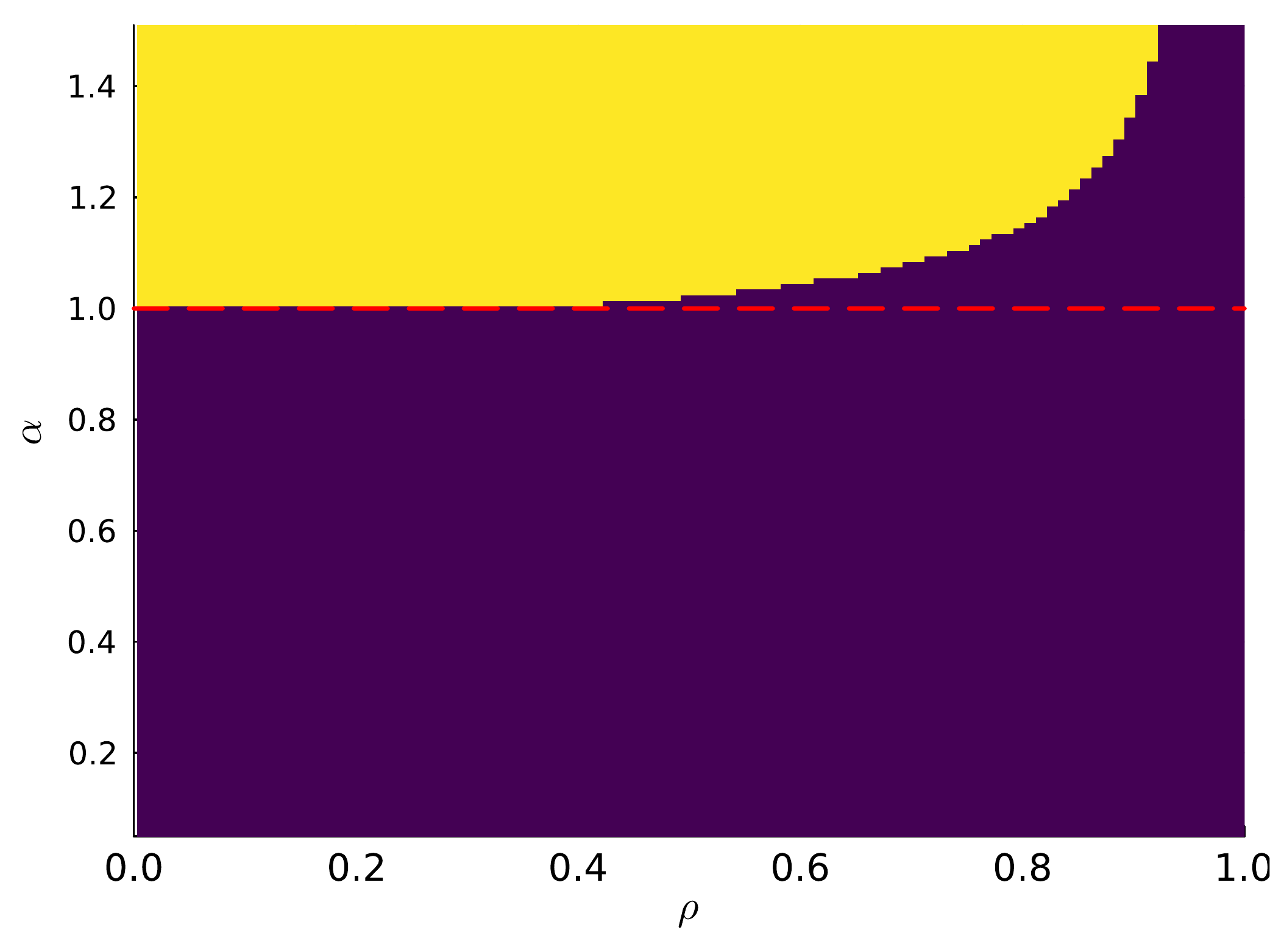}
        }
        \hfil
        \subfloat[Optimal $\ell_1$-ratio.]{%
            \label{fig:subfigB}
            \includegraphics[width=4.8cm]{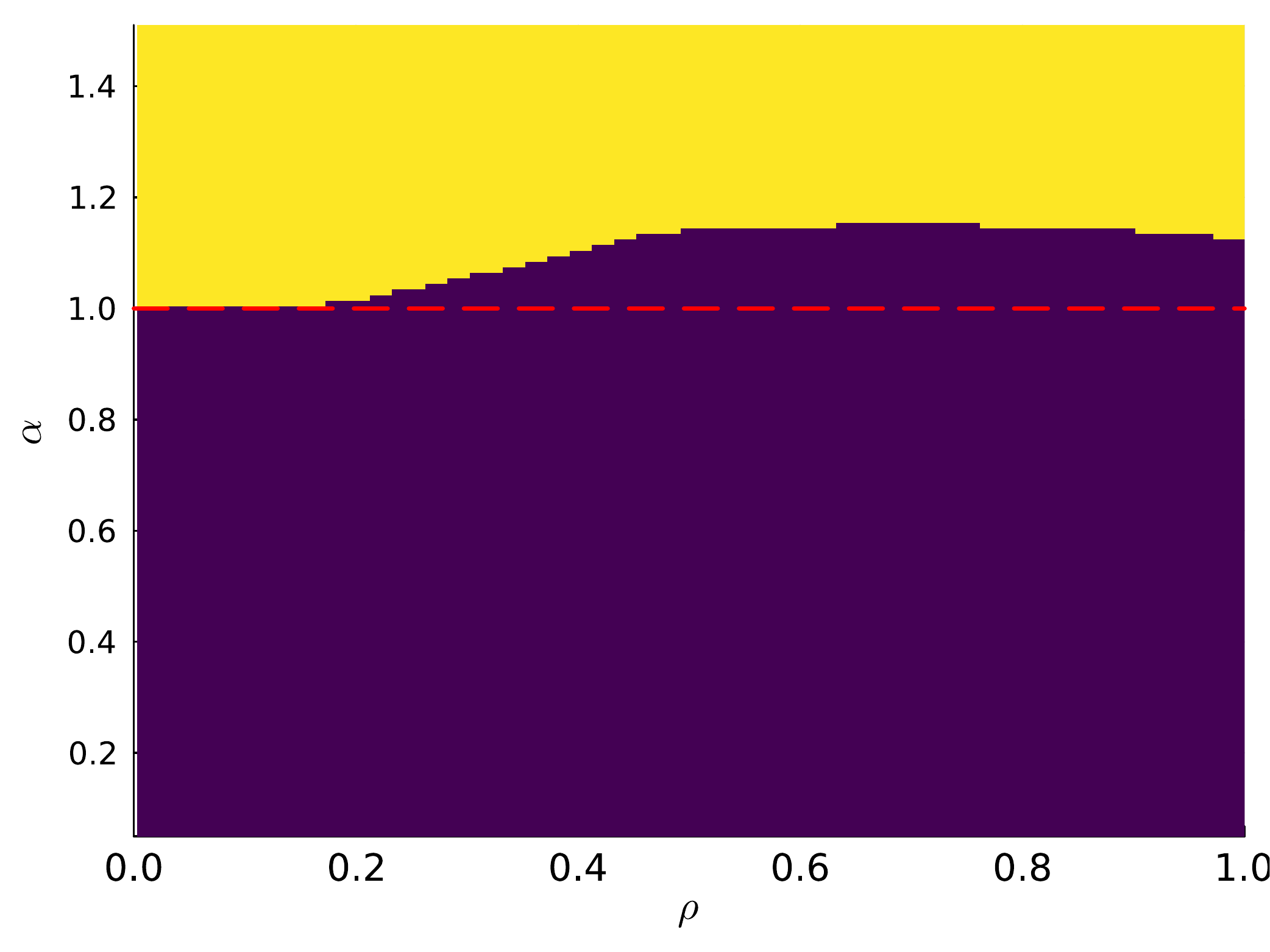}
        }
        \caption{The typical number of unique samples in each bootstrap sample $\alpha(1-e^{-\mu_B^\star})$ scaled by $N$ is visualized. In the purple region, $\alpha(1-e^{-\mu_B^\star})<1$, and in the yellow region, $\alpha(1-e^{-\mu_B^\star})\ge1$. The red dashed line shows $\alpha=1$.  $\Delta=0.15$.}
        \label{fig: is overparametrized}
    \end{figure*}

\subsection{Phase transition to an ensemble of interpolators}
\label{subsection: phase transition}
Fig. \ref{fig: optimal regularization} shows the optimal regularization strength $\lambda^\star$ for $\hat{\sigma}^2$. Panels (a) and (b) show the results when the $\ell_1$-ratio is fixed, and panel (c) shows when the $\ell_1$-ratio is optimized. In all cases, it is clear that a phase transition has occurred in which $\lambda^\ast$ drops to an infinitesimally small value (although for numerical reasons, $\lambda$ is constrained to exceed $10^{-7}$) as the measurement ratio $\alpha$ decreases or $\rho$ increases. Moreover, Fig. \ref{fig: is overparametrized} shows $\alpha(1-\mu_B^\star)$, the typical number of unique data points in each bootstrap sample scaled by $N$: $\lim_{M, N\to\infty}(M\mu_B)/N$. From Fig. \ref{fig: is overparametrized}, it is clear that the typical number of unique data points is always smaller than $1$ in the region where $\lambda^\star \simeq +0$. This holds, even if the measurement ratio $\alpha>1$. Thus, when $\lambda^\star\simeq+0$, the elastic net estimator in \eqref{eq: elastic net single realization of c} becomes the minimum elastic net norm estimator as 
\vspace{-1truemm}
\begin{multline}
    \hat{\bm{w}}^\ast = \min_{\bm{w}\in\mathbb{R}^N} \sum_{i=1}^N \gamma|w_i| + \frac{1-\gamma}{2}w_i^2,
    \, {\rm subject\,to}
    \\
     \quad \1[c_\mu>0](y_\mu - \bm{x}_\mu^\top{w}) = 0, \, \mu=1,2,\dots,M,
\end{multline}
which is commonly known as minimum elastic net norm \emph{interpolator} in machine learning \cite{muthukumar2020harmless, bartlett2020benign, hastie2022surprises}. These suggest that when elastic net regularization cannot determine an appropriate sparse structure of $\bm{w}_0$, it is better to use an over-parameterized setting in which the number of unique data points of each bootstrap data is smaller than the dimension of $\bm{w}_0$ and use an ensemble of interpolators.

\section{Summary and discussion}
\label{sec: summary and discussion}
In this study, we investigated the behavior of the bootstrap-averaged unbiased estimator of GAMP using AMPR and its SE. We found that the bootstrap averaging procedure can effectively reduce the variance of the unbiased estimator when the actual data generation process is inconsistent with the sparsity assumption of the regularization and the data size is insufficient. We also found a phase transition where the regularization strength drops to infinitesimally small values by decreasing the measurement ratio $\alpha$ or increasing $\rho$. 

Although increasing the variance of weak learners is key to the success of ensemble learning \cite{krogh1997statistical, sollich1995learning, kuncheva2003measures}, the phase transition to an ensemble of interpolators may be unexpected. Investigating whether similar phase transitions occur in other more sophisticated machine-learning models, such as neural networks, would be an interesting future direction. 

We also remark that the landscape in $(\mu_B, \lambda)$-space may not be convex in general. Thus, the uniqueness of the optimizer should be investigated in the future, paying attention to the relationship with the implicit regularization \cite{lejeune2020implicit,du2023subsample, yao2021minipatch}.

On the technical side, the key to this study was the precise performance characterization of the averaged estimator by AMPR, which was developed by combining GAMP and the replica method of statistical physics \cite{mezard2009information, montanari2022short}. Such a combination of approximate inference algorithms and the replica method has been used to develop approximate computation algorithms \cite{obuchi2019semi, takahashi2019replicated, takahashi2020semi, malzahn2003approximate, malzahn2003statistical, NIPS2003_2c6ae45a, opper2005approximate} and has not been applied to precise performance analysis of ensemble methods. It would be an interesting direction to try similar performance analysis for other bootstrap methods or ensemble learning.

\section*{Acknowledgement}
This study was supported by JSPS KAKENHI Grant Number 21K21310 and 17H00764.

\bibliographystyle{IEEEtran}
\bibliography{main}

\end{document}